\documentclass[11pt,a4j]{article}
\usepackage{amsmath,amssymb}
\usepackage{bm}
\usepackage{graphicx}
\usepackage{verbatim}
\usepackage{wrapfig}
\usepackage{ascmac}
\usepackage{relsize,amsmath,amssymb}
\usepackage{subcaption}
\usepackage{float}
\usepackage{mathrsfs}
\usepackage{fancybox}
\usepackage{here}
\usepackage{cite}
\newcommand\la{\langle}
\newcommand\ra{\rangle}

\newcommand\beq{\begin{eqnarray}}
\newcommand\eeq{\end{eqnarray}}
\newcommand\beqs{\begin{eqnarray*}}
\newcommand\eeqs{\end{eqnarray*}}

\def\Pslash{\rlap/{\mkern-1mu P}}

\def\Sslash{\rlap/{\mkern-1mu S}}
\def\pslash{\rlap/{\mkern-1mu p}}

\def\nn{\nonumber}
\def\del{\partial}
\def\epal{\epsilon^{\alpha S_\perp{w} P_h}}

\def\zh{\hat{z}}

\def\xh{\hat{x}}

\def\hz{\hat{z}}
\def\hx{\hat{x}}

\def\qt{q_T}

\def\3Tb{3\bar{T}}

\allowdisplaybreaks[1]

\begin{document}
\begin{flushright}
\today
\end{flushright}
\vspace*{5mm}
\begin{center}
{\Large \bf Transverse Polarization of Hyperons Produced
in Semi-Inclusive \\[10pt] Deep Inelastic Scattering}
\vspace{1.5cm}\\
{\sc Yuji Koike$^1$, Kazuki Takada$^2$, Sumire Usui$^3$,  
Kenta Yabe$^2$ and Shinsuke Yoshida$^{4,5}$}
\\[0.7cm]
\vspace*{0.1cm}
{\it $^1$ Department of Physics, Niigata University, Ikarashi 2-no-cho, Niigata 950-2181, Japan}

\vspace{0.2cm}

{\it $^2$ Graduate School of Science and Technology, Niigata University,
Ikarashi 2-no-cho, Niigata 950-2181, Japan}

\vspace{0.2cm}

{\it $^3$ Enterprise Business Unit, NEC Corporation, Shiba 5, Minato-ku, Tokyo 108-8001, Japan}

\vspace{0.2cm}

{\it $^4$ Guangdong Provincial Key Laboratory of Nuclear Science, 
Institute of Quantum Matter,\\
South China Normal University, Guangzhou 510006, China}

\vspace{0.2cm}

{\it $^5$ Guangdong-Hong Kong Joint Laboratory of Quantum Matter, 
Southern Nuclear Science Computing Center, 
South China Normal University, Guangzhou 510006, China}
\\[3cm]

{\large \bf Abstract} \end{center}
We study the transverse polarization of hyperons produced in semi-inclusive 
deep inelastic scattering, $ep\to e\Lambda^\uparrow X$,  
in the framework of the collinear twist-3 factorization.
The cross section from the twist-3 distribution functions and the twist-3
quark fragmentation functions is computed in the leading order with respect to the QCD 
coupling constant.  The constraint relations among the twist-3 FFs are taken into account
to simplify the formula.  The formula is relevant to 
large-$P_T$ hyperon production in the future Electron-Ion-Collider experiment.

\newpage

\section{Introduction}
Transverse polarization of hyperons produced in unpolarized collisions
have been observed in many high-energy inclusive processes such as 
$pp\to\Lambda^\uparrow X$, 
$e^+e^-\to\Lambda^\uparrow X$ and $ep\to e\Lambda^\uparrow X$\footnote{Throughout 
this paper we collectively denote spin-1/2 hyperons as $\Lambda$.}.  
This phenomenon is an example of transverse single spin asymmetries (SSAs)
in which only one particle appearing in the process is transversely polarized.  
The SSAs associated with the initial state 
spin such as $p^\uparrow p\to hX$ and
$ep^\uparrow \to ehX$ ($h=\pi,K,\eta,...$) have also been observed.  
The parton model or 
perturbative QCD at twist-2 level fails to produce large SSAs\cite{Kane:1978nd}, and 
its description in terms of QCD has been a challenge in QCD spin physics.  
In the collinear factorization of perturbative QCD, the SSAs occur as a twist-3 observable,
which reflects multi-parton correlations either in the initial nucleon or in the fragmentation 
processes\cite{Efremov:1981sh,Qiu:1991wg,Qiu:1998ia}.
Derivation of the twist-3 cross sections for SSAs required lots of technical development.  
By now the collinear twist-3 formalism for all kinds of the 
twist-3 distribution functions (DFs) and the fragmentation functions (FFs) has 
been well established in the
leading order (LO) with respect to the QCD coupling 
constant\cite{Eguchi:2006qz,Eguchi:2006mc,Ji:2006vf,Kouvaris:2006zy,Koike:2007rq,Koike:2006qv,
Koike:2007dg,Koike:2009ge,Yuan:2009dw,Kang:2010zzb,Beppu:2010qn,Koike:2011ns,Kanazawa:2013uia,
Hatta:2013wsa,Metz:2012ct,Koike:2021awj}.  
Those functions
are, in general, not independent from each other but obey some constraint
relations based on the operator identities. 
The complete set of those relations have been also derived\cite{Kanazawa:2015ajw,Koike:2019zxc},
which are crucial to obtain the frame-independent expressions for the twist-3 
cross sections\cite{Kanazawa:2014tda,Kanazawa:2015ajw,Koike:2019zxc}.  
There have been also some attempts to extend the twist-3 calculation to 
the next-to-leading order level\cite{Gamberg:2018fwy,Benic:2019zvg,Benic:2021gya}.  

In this paper, we study the transverse polarization of hyperons produced in semi-inclusive deep inelastic 
scattering (SIDIS), $ep\to e\Lambda^\uparrow X$,
in the LO collinear twist-3 factorization.  
As for the case of $pp\to \Lambda^\uparrow X$\cite{Kanazawa:2001a,Zhou:2008,Koike:2015zya,
Koike:2017fxr,Koike:2021awj}, 
two kinds of the twist-3 cross sections contribute.  
One is from the twist-3 DF in unpolarized proton combined with
the ``transversity'' FF for the polarized $\Lambda$.   
The other one is from the
twist-3 polarized FFs for $\Lambda$ combined with the unpolarized parton DFs in the proton.
In the latter contribution, both twist-3 quark FFs\cite{Koike:2017fxr}
and twist-3 gluon FFs\cite{Koike:2021awj} contribute due to the
chiral-even nature of the FFs and DFs.  
Here we report the LO twist-3 cross section formula for $ep\to e\Lambda^\uparrow X$
from the twist-3 DFs and the twist-3 quark FFs. 
This is relevant for the large-$P_T$ polarized hyperon production in the
future Electron-Ion-Collider (EIC) experiment.  
The contribution from
twist-3 purely gluonic FFs will be reported in a separate publication\cite{ikarashi2022}.

The remainder of this paper is organized as follows:
In section 2, we summarize the twist-3 DFs and FFs relevant to 
the present study.  
In section 3, after summarizing the kinematics of $ep\to e\Lambda^\uparrow X$ (Sec 3.1),
we present the cross section from the
twist-3 DFs (Sec. 3.2) and from the twist-3 quark FFs (Sec. 3.3).  
Section 4 is devoted to a brief summary of the present study.


\section{Twist-3 quark DFs and FFs}

In this section, we summarize the twist-3 distribution function (DF)
and the fragmentation functions (FF), which are necessary to calculate
the polarized cross section for $ep\to e\Lambda^\uparrow X$.  
For the twist-3 distribution, we need only one function
$E_F(x_1,x_2)$ in an unpolarized proton, which
is defined from the quark-gluon correlation function as\cite{Kanazawa:2000hz}
\beq
M_F^\alpha (x_1,x_2)&=&
\int {d\lambda\over 2\pi}\int {d\mu\over 2\pi}e^{i\lambda x_1}e^{i\mu(x_2-x_1)}
\la p| \bar{\psi}_j(0)[0,\mu n]gF^{\alpha n}(\mu n)  [\mu n, \lambda n] \psi_i(\lambda n)|p\ra\nn\\
&=&-{M_N\over 4}\epsilon^{\alpha\beta np}(\gamma_5\gamma_\beta \pslash )_{ij}E_F(x_1,x_2)
+\cdots, 
\label{tw3distribution}
\eeq
where $M_N$ is the nucleon mass, 
$| p\ra$ is the nucleon state with momentum $p$ which can be regarded as lightlike,  
$n$ is another lightlike vector
satisfying $p\cdot n=1$, and $i,\ j$ denotes the spinor indices.  
We use the convention for the $\epsilon$-tensor as 
$\epsilon^{0123}=1$
and the notation 
$\epsilon^{\alpha\beta np}\equiv \epsilon^{\alpha\beta\mu\nu}n_\mu p_\nu$ is used. 
$[\mu n, \lambda n] ={\rm P}\,{\rm exp}\left[ig\int^{\mu}_{\lambda} d\tau\, n\cdot A(\tau n)\right]$ 
is the gauge link operator, which makes the correlation function color gauge invariant.   
From $P,\ T$-invariance, one has $E_F(x_1,x_2)=E_F(x_2,x_1)$.  
The support of $E_F(x_1,x_2)$ is $|x_{1,2}|<1$ and $|x_1-x_2|<1$.  
The anti-quark or ``charge-conjugated'' distribution $\bar{E}_F(x_1,x_2)$ for $E_F$ is defined
by $\psi\to C\bar{\psi}^T$, $\bar{\psi}\to -\psi^T C^{-1}$ 
($C$ is the charge conjugation matrix) and 
$F_{\alpha\beta}\to - F_{\alpha\beta}^T$ in (\ref{tw3distribution}) 
and it satisfies the relation $\bar{E}_F(x_1,x_2)=E_F(-x_2,-x_1)$.

We need several kinds of FFs, which are summarized below using the notation in 
\cite{Kanazawa:2015ajw}.
The simplest ones are defined from
the lightcone correlation functions of quark fields: 
\begin{eqnarray}
\Delta_{ij}(z)&=&{1\over N}\sum_X\!\int \!\!
\frac{d\lambda}{2\pi} e^{-i{\lambda\over z}}\la 0|[\infty w, 0]\psi_i(0)| 
h(P_h,S_\perp)X\ra\la h(P_h,S_\perp)X|\bar{\psi}_j(\lambda w)
[\lambda w,\infty w]| 0\ra\nn\\
&=&\left(\gamma_5\Sslash_\perp\frac{\Pslash_h}{z}\right)_{ij} H_1(z)+M_h
\epal(\gamma_\alpha)_{ij}\frac{D_T(z)}{z}+\,M_h(\gamma_5\Sslash_\perp)_{ij}\frac{G_T(z)}{z}+\cdots.
\label{qFraI}
\end{eqnarray}
where $|h(P_h,S_\perp)\ra$ denotes the hyperon state with mass $M_h$, 
momentum $P_h$
and the transverse spin vector $S_\perp$ normalized as $S_\perp^2=-1$.  
Since we are interested in the twist-3 cross section,
we treat $P_h$ as lightlike, and $w$ is another lightlike vector satisfying $P_h\cdot w=1$.  
$H_1(z)$ is the twist-2 transversity FF, 
$D_T(z)$ and $G_T(z)$ are twist-3 and are called {\it intrinsic} twist-3 FFs.  
${D_T(z)}$ is naively $T$-odd, while $G_T(z)$ is naively $T$-even.  
In (\ref{qFraI}), gauge link operator 
$[\lambda w,\infty w]\equiv 
{\rm P}{\rm exp}\left[ig\int^\lambda_\infty d\tau\, w\cdot A(\tau w)\right],$
is inserted, which makes the correlation function gauge invariant.

The next one is the twist-3 {\it kinematical} FFs which are defined as
\begin{eqnarray}
\Delta_{\del ij}^\alpha(z)
&=&{1\over N}\sum_X\!\!\int \!\!\frac{d\lambda}{2\pi} 
e^{-i{\lambda\over z}}\la 0|[\infty w,0]\psi_i(0)| h(P_h,S_\perp)X
\ra \la h(P_h,S_\perp)X|\bar{\psi}_j(\lambda w)
[\lambda w,\infty w]| 0\ra\overleftarrow{\del}^\alpha\nn\\
&=&-iM_h\epal(\Pslash_h)_{ij}\frac{D_{1T}^{\perp(1)}(z)}{z}+\, iM_hS_\perp^\alpha
(\gamma_5\Pslash_h)_{ij}\frac{G_{1T}^{\perp(1)}(z)}{z}+\cdots, 
\label{qFraK}
\end{eqnarray}
where each FF is defined to be real, and they are related to $k_T^2/M_h^2$-moment of
the transvers-mometum-dependent (TMD) FFs\cite{Mulders:1995dh}.  
$G_{1T}^{\perp(1)}(z)$ is naively $T$-even, while
$D_{1T}^{\perp(1)}(z)$ is naively $T$-odd and 
contributes to the hyperon polarization.

Next we introduce the twist-3 {\it dynamical} FFs
which are defined from the
three parton correlation function as:
\begin{eqnarray}
&&\hspace{-0.5cm}
\Delta_{F ij}^\alpha(z,z_1)\nn\\[3pt]
&&\quad={1\over N}\sum_X\!\int \!\!\frac{d\lambda}{2\pi}\! \int \!\!\frac{d\mu}{2\pi} 
e^{-i{\lambda\over z_1}}e^{-i\mu({1\over z}-{1\over z_1})}\la 0|\psi_i(0)| h(P_h, S_\perp)X\ra\la h(P_h,S_\perp)X|\bar{\psi}_j(\lambda w)gF^{\alpha w}(\mu w)| 0\ra\nn\\[5pt]
&&\quad=M_h \epal(\Pslash_h)_{ij}\frac{\widehat{D}_{FT}^\ast(z,z_1)}{z}
-\,iM_h 
S_\perp^\alpha(\gamma_5\Pslash_h)_{ij}\frac{\widehat{G}_{FT}^\ast(z,z_1)}{z}+\cdots, 
\label{qFraD}
\end{eqnarray}
where the gauge link is suppressed for simplicity.  The dynamical FFs
$\widehat{D}_{FT}(z,z_1)$ and 
$\widehat{G}_{FT}(z,z_1)$ are complex functions and their complex conjugates are
defined in (\ref{qFraD}).  
The real parts of these functions are naively $T$-even, while the
imaginary parts are naively $T$-odd and
contribute to the hyperon polarization.  
Replacing $gF^{w\alpha}(\mu w)$ by the covariant derivative 
$D^\alpha(\mu w)=\partial^\alpha -ig A^\alpha(\mu w)$, one can define another set of the
twist-3 FFs, $\widehat{D}_{DT}(z,z_1)$ and 
$\widehat{G}_{DT}(z,z_1)$, by the same tensor decomposition as above.  
But they can be related to
the above functions\cite{Metz:2012ct,Kanazawa:2013uia}:
\begin{eqnarray}
{\Im} \, \widehat{D}_{DT}(z,z_1) & = & P \frac{1}{1/z - 1/z_1} {\Im} \, \widehat{D}_{FT}(z,z_1) 
-\delta\Big( \frac{1}{z} - \frac{1}{z_1} \Big) D_{1T}^{\perp (1)}(z) ,
\label{e:DD_DF} \\ 
{\Im} \, \widehat{G}_{DT}(z,z_1) & = & P \frac{1}{1/z - 1/z_1} {\Im} \, \widehat{G}_{FT}(z,z_1) .
\label{e:GD_GF}
\end{eqnarray}

Three kinds of twist-3 FFs (\ref{qFraI}), (\ref{qFraK}) and (\ref{qFraD}) are not independent, 
but are subject to the EOM relation and the Lorentz invariance relations (LIRs).
Here we quote those relations from \cite{Kanazawa:2015ajw}.  
The EOM relation involving the naively $T$-odd FFs is given by
\begin{eqnarray}
\int_z^\infty\frac{dz_1}{z_1^2}\frac{1}{1/z-1/z_1}\Bigl(\Im\widehat{D}_{FT}(z,z_1)
-\Im\widehat{G}_{FT}(z,z_1)\Bigr)
=\frac{D_T(z)}{z}+D_{1T}^{\perp(1)}(z),
\label{qFra_EOM}
\end{eqnarray}
and the LIR reads
\begin{eqnarray}
-{2\over z}\!\int_z^\infty\!\frac{dz_1}{z_1^2}\frac{{\Im}\widehat{D}_{FT}(z,z_1)}{(1/z_1-1/z)^2} 
=\frac{D_T(z)}{z}+\frac{d\!\left(\!D_{1T}^{\perp(1)}(z)/z\!\right)}{d(1/z)}. 
\label{qFra_LIR}
\end{eqnarray}
It has been shown that
these relations are crucial to guarantee the gauge invariance and the frame independence of
twist-3 cross sections for various processes\cite{Kanazawa:2015ajw,
Kanazawa:2014tda,Koike:2017fxr}.  

In addition to those in  (\ref{qFraD}), there are another type of dynamical FFs defined from
the matrix elements like $\sim\la 0|gF_a^{\alpha w}|hX\ra \la hX| \bar{\psi} t^a\psi|0\ra$
with $t^a$ the generator of color SU(3).
They are, however, related to the purely gluonic twist-3 FFs by the EOM relations and 
the LIRs as was shown in \cite{Koike:2019zxc}.  
It's been also shown that the combination of
the contributions from the twist-3 purely gluonic FFs and 
these dynamical FFs gives the gauge and frame independent cross section
for $pp\to\Lambda^\uparrow X$ thanks to the LIR s and the EOM relations
\cite{Koike:2021awj}.  Therefore we will discuss those contributions 
together in a separate publication
\cite{ikarashi2022}.

\section{Twist-3 cross section for $ep\to e\Lambda^\uparrow X$}

\subsection{Kinematics}
Here we summarize the kinematics for 
\begin{eqnarray}
e(\ell)+p(p)\to e(\ell')+\Lambda^\uparrow(P_h,S_\perp)+X,
\label{sidishyperon}
\end{eqnarray}
where $\ell$,  $\ell'$, $p$ and $P_h$ are the momenta of each particle, and $S_\perp$
is the transverse spin vector of the produced $\Lambda^\uparrow$.  
To derive the cross section for this process, we define the
following five Lorentz invariants: 
\beq
S_{ep}&=&(p+\ell)^2,\nn\\
x_{bj}&=&\frac{Q^2}{2p\cdot q},\nn\\
Q^2&=&-q^2=-(\ell-\ell')^2,\nn\\
z_f&=&\frac{p\cdot P_h}{p\cdot q},\nn\\
q_T&=&\sqrt{-q_t^2},
\label{kinvar}
\eeq
where the space-like four momentum $q_t$ is defined by
\beq
q_t^{\mu}=q^{\mu}-\frac{P_h\cdot q}{p\cdot P_h}p^{\mu}-\frac{p\cdot q}{p\cdot P_h}P_h^{\mu},
\eeq
which satisfies $q_t\cdot p=q_t\cdot P_h=0$.  
\begin{figure}[H]
	\centering
	\includegraphics[keepaspectratio,scale=0.53]{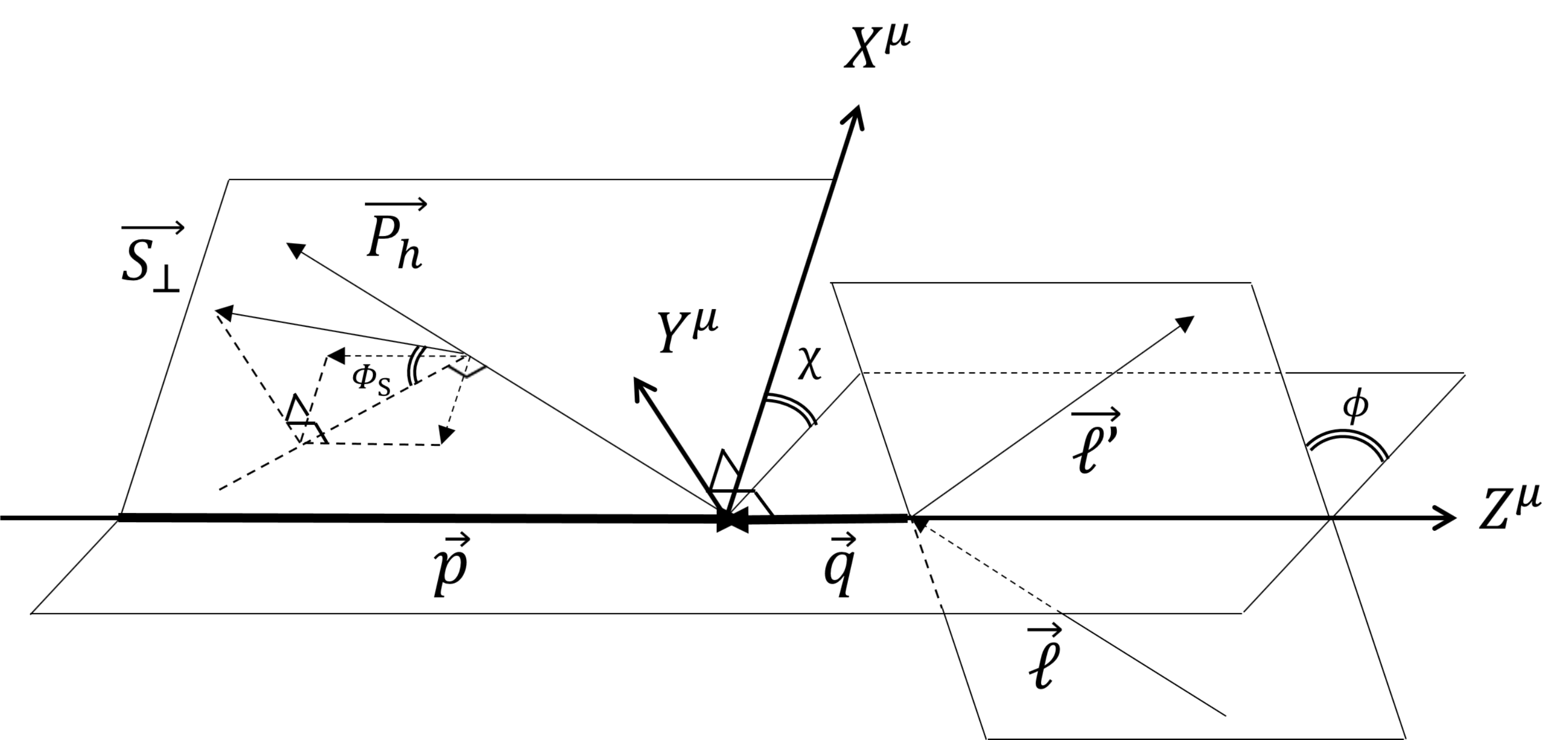}
	\caption{Hadron frame in which $\vec{q}$ and $\vec{p}$ are collinear. 
	The angles $\phi$ and $\chi$ are, respectively, the
	azimuthal angles for the lepton plane and the hadron plane (measured from a 
	certain reference plane), 
	and $\Phi_S$ is the azimuthal angle of the transverse spin vector of 
	$\Lambda^\uparrow$, $\vec{S}_\perp$, measured from the hadron plane. }
	\label{hadronframe}
\end{figure}
We work in the hadron frame
\cite{Meng:1991da}, in which the momenta
of the virtual photon and the initial proton are collinear as shown in Fig. \ref{hadronframe}.  
In this frame those momenta take
\begin{eqnarray}
q^\mu&=&(q^0,\overrightarrow{q})=(0,0,0,-Q),\\
p^{\mu}&=&\left(\frac{Q}{2x_{bj}},0,0,\frac{Q}{2x_{bj}}\right).  
\end{eqnarray}
We define the azimuthal angles of the hadron plane and the lepton plane 
as $\chi$ and $\phi$, 
respectively.   
With these angles, $P_h^\mu$ can be written as
\beq
P_h^{\mu}=\frac{z_fQ}{2}\left(1+\frac{q_T^2}{Q^2},\frac{2q_T}{Q}{\rm cos}\chi,\frac{2q_T}{Q}{\rm sin}\chi,-1+\frac{q_T^2}{Q^2}\right),
\eeq
which implies the transverse momentum of $\Lambda^\uparrow$ is 
$P_{hT}=z_f q_T$.  
For this $P_h^{\mu}$, $w^\mu$ in (\ref{qFraI}) takes the following form: 
\begin{eqnarray}
w^{\mu}=\frac{1}{z_fQ(1+q_T^2/Q^2)^2}\left(1+\frac{q_T^2}{Q^2},-\frac{2q_T}{Q}{\rm cos}\chi,-\frac{2q_T}{Q}{\rm sin}\chi,1-\frac{q_T^2}{Q^2}\right).  
\end{eqnarray}
The initial and the scattered lepton momenta are, respectively, given by
\begin{eqnarray}
\ell^\mu&=&\frac{Q}{2}({\rm cosh}\psi,{\rm sinh}\psi {\rm cos}\phi,{\rm sinh}
\psi {\rm sin}\phi,-1),\\
\ell'^\mu&=&\frac{Q}{2}({\rm cosh}\psi,{\rm sinh}\psi {\rm cos}\phi,{\rm sinh}
\psi {\rm sin}\phi,1), 
\end{eqnarray}
with
\beq
\cosh\psi=\frac{2x_{bj}S_{ep}}{Q^2}-1.
\eeq

In order to calculate the cross section, we introduce the following
four vectors orthogonal to each other\cite{Meng:1991da},
\begin{eqnarray}
&&T^\mu=\frac{1}{Q}(q^\mu+2x_{bj}p^\mu)=(1,0,0,0),\\
&&X^\mu=\frac{1}{q_T}\left\{\frac{P_h^\mu}{z_f}-q^\mu-(1+\frac{q_T^2}{Q^2})x_{bj}p^\mu\right\}=(0,{\rm cos}\chi,{\rm sin}\chi,0),\\
&&Z^\mu=-\frac{q^\mu}{Q}=(0,0,0,1),\\
&&Y^\mu=\epsilon^{\mu\nu\rho\sigma}T_\nu X_\rho Z_\sigma
=(0,-{\rm sin}\chi,{\rm cos}\chi,0).
\end{eqnarray}
The polar angle $\theta$ of $\vec{P_h}$ measured from the $Z$-axis can be written as
\begin{eqnarray}
{\rm cos}\theta=\frac{P_{hz}}{|\vec{P_h}|}=\frac{q_T^2-Q^2}{q_T^2+Q^2}, \\
{\rm sin}\theta=\frac{P_{hT}}{|\vec{P_h}|}=\frac{2q_TQ}{q_T^2+Q^2},
\end{eqnarray}
where $P_{hT}=\sqrt{P_{hx}^2+P_{hy}^2}$.  
The transverse spin vector $\vec{S}_\perp$ of the hyperon resides in the plane
which is orthogonal to $\vec{P}_h$, and we define the azimuthal angle of $\vec{S}_\perp$
measured from the hadron plane around $\vec{P}_h$ as $\Phi_s$
(See Fig. \ref{hadronframe}).
Then $S_\perp^\mu$ can be written as
\beq
S_\perp^\mu=\cos\theta \cos\Phi_S\, X^\mu+\sin\Phi_S \,Y^\mu
-\sin\theta\cos\Phi_S\, Z^\mu.  
\eeq

The polarized cross section for (\ref{sidishyperon}) can be written as
\beq
d\Delta\sigma=\frac{1}{2S_{ep}}\frac{d^{3}\overrightarrow{P_{h}}}{(2\pi)^{3}2P^{0}_{h}}
\frac{d^{3}\overrightarrow{\ell '}}{(2\pi)^{3}2\ell'^{0}}
\frac{e^{4}}{q^{4}}L^{\mu \nu}\left(l ,l' \right)W_{\mu \nu}\left(p,q,P_{h}\right), 
\label{dsigma}
\eeq
where $W_{\mu\nu}$ is the hadronic tensor and 
$L^{\mu\nu}$ is the unpolarized leptonic tensor defined by
\beq
L^{\mu\nu}(\ell,\ell')=2(\ell^\mu \ell'^\nu+\ell^\nu \ell'^{\mu})-Q^2g^{\mu\nu}.
\label{letontensor}
\eeq
Using the kinematic variables introduced in (\ref{kinvar}), the differential cross section
can be written as
\beq
\frac{d^{6}\Delta\sigma}{dx_{bj}dQ^{2}dz_{f}dq^{2}_{T}d\phi d\chi}  =
\frac{\alpha ^{2}_{em}z_f}{128\pi^4 S^{2}_{ep} x^{2}_{bj} Q^{2}}
L^{\mu\nu}(\ell,\ell')W_{\mu\nu}(p,q,P_h), 
\label{Xsec}
\eeq
where $\alpha_{em}=e^2/(4\pi)$ is the fine structure constant in QED.  
The hadronic tensor $W_{\mu\nu}$ satisfies the current conservation
$q^\mu W_{\mu\nu}=q^\nu W_{\mu\nu}=0$ and can be expanded by the
following six tensors, 
$\mathscr{V}_k^{\mu\nu}$\,\cite{Meng:1991da}:
\begin{eqnarray}
&&\mathscr{V}_1^{\mu\nu}=X^\mu X^\nu+Y^\mu Y^\nu,\nn\\
&&\mathscr{V}_2^{\mu\nu}=g^{\mu\nu}+Z^\mu Z^\nu,\nn\\
&&\mathscr{V}_3^{\mu\nu}=T^\mu X^\nu+X^\mu T^\nu,\nn\\
&&\mathscr{V}_4^{\mu\nu}=X^\mu X^\nu-Y^\mu Y^\nu,\\
&&\mathscr{V}_8^{\mu\nu}=T^\mu Y^\nu+Y^\mu T^\nu,\nn\\
&&\mathscr{V}_9^{\mu\nu}=X^\mu Y^\nu+Y^\mu X^\nu.\nn
\end{eqnarray}
In order to 
calculate $L^{\mu\nu} W_{\mu\nu}$ in (\ref{dsigma}), 
we introduce the inverse tensors 
$\widetilde{\mathscr{V}}_k^{\mu\nu}$ for $\mathscr{V}_k^{\mu\nu}$: 
\begin{eqnarray}
&&\widetilde{\mathscr{V}}_1^{\mu\nu}=
\frac{1}{2}(2T^\mu T^\nu+X^\mu X^\nu+Y^\mu Y^\nu),\nn\\
&&\widetilde{\mathscr{V}}_2^{\mu\nu}=T^\mu T^\nu,\nn\\
&&\widetilde{\mathscr{V}}_3^{\mu\nu}=-\frac{1}{2}(T^\mu X^\nu+X^\mu T^\nu),\nn\\
&&\widetilde{\mathscr{V}}_4^{\mu\nu}=\frac{1}{2}(X^\mu X^\nu-Y^\mu Y^\nu),\\
&&\widetilde{\mathscr{V}}_8^{\mu\nu}=-\frac{1}{2}(T^\mu Y^\nu+Y^\mu T^\nu),\nn\\
&&\widetilde{\mathscr{V}}_9^{\mu\nu}=\frac{1}{2}(X^\mu Y^\nu+Y^\mu X^\nu).\nn
\end{eqnarray}
With these $\mathscr{V}_k^{\mu\nu}$ and $\widetilde{\mathscr{V}}_k^{\mu\nu}$, one obtains
$L^{\mu\nu} W_{\mu\nu}$ as
\begin{eqnarray}
L^{\mu\nu}W_{\mu\nu}=\sum_{k=1,\cdots, 9}[L_{\mu\nu}\mathscr{V}_k^{\mu\nu}][W_{\rho\sigma}\widetilde{\mathscr{V}}_{k}^{\rho\sigma}]=Q^2\sum_{k=1,\cdots, 9}\mathcal{A}_k(\phi-\chi)[W_{\rho\sigma}\widetilde{\mathscr{V}}_k^{\rho\sigma}],
\label{LWproduct}
\end{eqnarray}
where $\mathcal{A}_k(\varphi)$ ($\varphi\equiv\phi-\chi$, $k=1,\cdots, \, 4,\, 8,\, 9$) is defined by
\begin{eqnarray}
\mathcal{A}_{\it k}(\varphi)=L_{\mu\nu}\mathscr{V}_{\it k}^{\mu\nu}/Q^2, 
\end{eqnarray}
and they are calculated to be
\begin{eqnarray}
&&\mathcal{A}_1(\varphi)=1+ \cosh^2\psi,\nn\\
&&\mathcal{A}_2(\varphi)=-2,\nn\\
&&\mathcal{A}_3(\varphi)=-{\rm cos}\varphi \sinh 2\psi,\nn\\
&&\mathcal{A}_4(\varphi)={\rm cos}2\varphi \sinh^2\psi,\\
&&\mathcal{A}_8(\varphi)=-{\rm sin}\varphi \sinh 2\psi,\nn\\
&&\mathcal{A}_9(\varphi)={\rm sin}2\varphi \sinh^2\psi. \nn
\end{eqnarray}
Corresponding to $\mathcal{A}_k(\phi-\chi)$ ($k=1,\ 2\cdots 9$), 
the cross section for (\ref{sidishyperon}) consists of five components
with different dependences on the azimuthal angle $\phi-\chi$. 

\subsection{Contribution from unpolarized twist-3 
distribution and the transversity fragmentation function}

\begin{figure}[h]
\vspace{-0.5cm}
    \begin{center}
	\includegraphics[width=16cm]{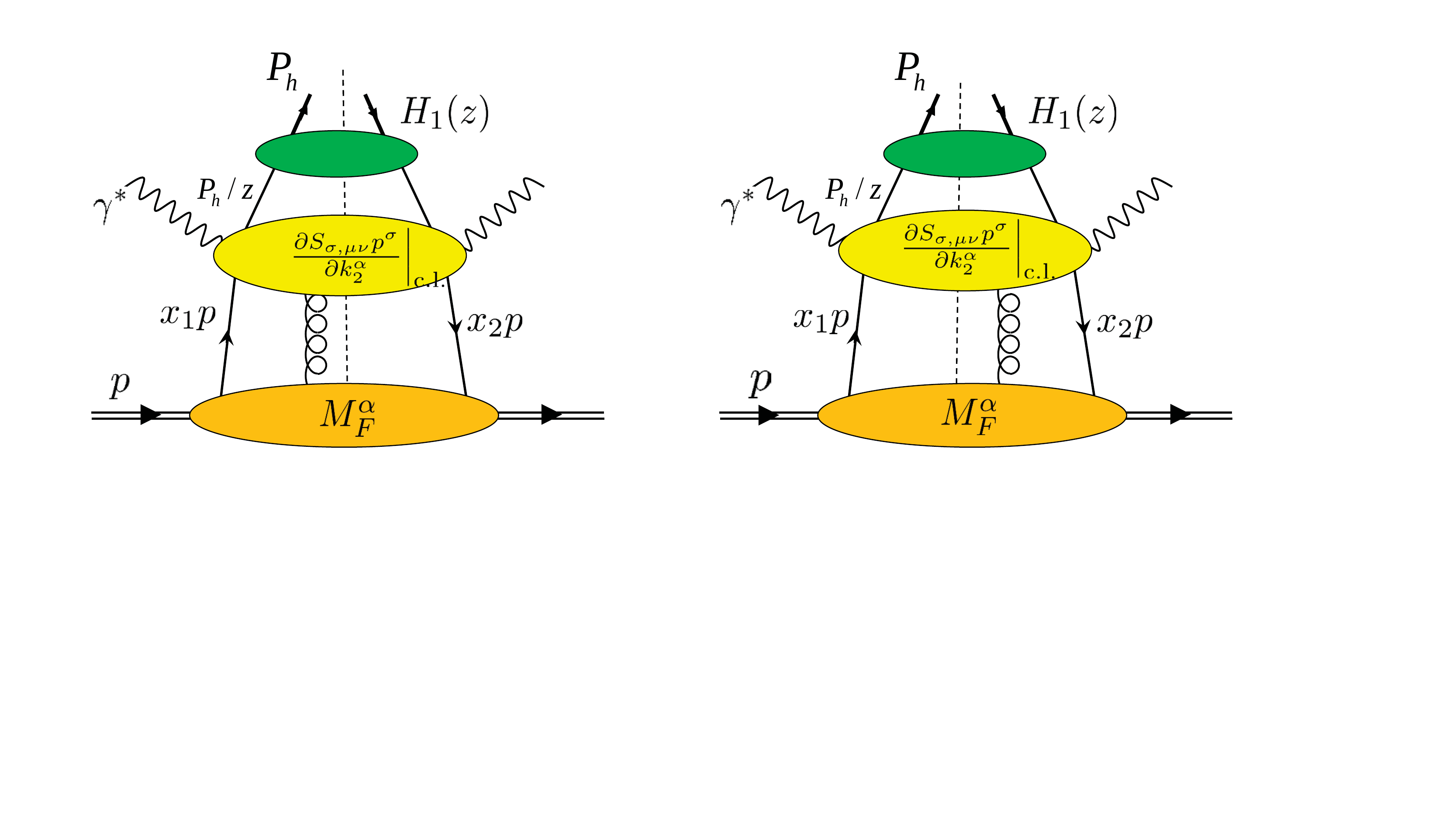}
    \end{center}
    \vspace{-4.0cm}
	\caption{Twist-3 distribution function contribution to $ep\to e\Lambda^\uparrow X$.
The top blob indicates the fragmentation correlator for $H_1(z)$, the bottom one represents
the correlation function for $E_F(x_1,x_2)$  and
the middle one corresponds to the partonic hard part. }
	\label{tw3distributionFig}
  \end{figure}

In this section we calculate the twist-3 cross section
for $ep\to e\Lambda^\uparrow X$ which arises
from the twist-3 DF in the nucleon.   
The method of the calculation is described in detail in
\cite{Eguchi:2006mc}, which developed the formalism for a similar twist-3
process
$ep^\uparrow \to e\pi X$.  
At leading order with respect to the QCD 
coupling constant, 
only the dynamical twist-3 DF in the nucleon
$E_F(x_1,x_2)$ defined in (\ref{tw3distribution}) contributes together with the 
transversity FF $H_1(z)$ for $\Lambda^\uparrow$, which is
schematically shown in 
Fig. \ref{tw3distributionFig}.  
Since the original calculation in \cite{Eguchi:2006mc} overlooked
some of the diagrams for the hard part\cite{Koike:2007dg,Koike:2009yb}, 
we shall also include those new diagrams below. 
The cross section
occurs as pole contributions from the hard part, which are classified into the
hard pole (HP), soft-gluon-pole (SGP) and 
soft-fermion-pole (SFP).  
For the SFP contribution,
the new type of diagrams found in \cite{Koike:2007dg}
cancel the contribution from the quark's SFP function $E_F(0,x)$ paired with
the quark's transversity FF arising from the SFP diagrams
considered in \cite{Eguchi:2006mc}.
Accordingly, 
in the present case, the SFP contribution occurs from
the left four diagrams in Fig. 1 of  \cite{Koike:2009yb}
in which the lower quark line (for anti-quark) crossing the final state cut fragments
into $\Lambda^\uparrow$, together with the diagrams obtained by reversing the arrows
on the quark lines. 
Therefore the SFP contribution
survives only for the anti-quark's transversity FF 
paired with
the quark's SFP DF or the quark's transversity FF
paired with the {\it anti-quark's} SFP DF\cite{Koike:2009yb}.  
We expect these contributions involving the FF or SFP
functions for an anti-quark should be much smaller compared with those from the SGP and HP ones.  
Hence we will not consider them below.  
It was also found in \cite{Koike:2009yb} that there are some other 
HP diagrams not considered in \cite{Eguchi:2006mc}.  
We will also include those new types of HP contributions below.  

Using (\ref{Xsec}) and (\ref{LWproduct}), and 
factorizing the twist-2 transversity FF $H_1(z)$
from the hadronic tensor,
one can write this cross section as
\beq
\frac{d^6\Delta\sigma^{\rm tw3-dist}}{dx_{bj}dQ^2dz_fdq_T^2
d\phi d\chi}=\frac{z_f \alpha_{em}^2}{128\pi^4 S_{ep}^2x_{bj}^2}
\sum^9_{k=1}\mathcal{A}_k(\phi-\chi)\int \frac{dz}{z^2}
H_1(z)\left[w_{\mu \nu}(p,q,P_h/z)\widetilde{\mathcal{V}}^{\mu\nu}_k\right], 
\label{EFH1contribution}
\eeq
where the summation over quark flavors as well as the factor associated with
the quark's fractional electric charge is omitted.  
Applying the formalism described in \cite{Eguchi:2006mc}, 
we obtain $w_{\mu\nu}$ in (\ref{EFH1contribution})
in terms of the gauge-invariant correlation function 
$M_F^{\alpha}(x_1,x_2)$ in (\ref{tw3distribution}) as
\beq
w_{\mu\nu}(p,q,P_h/z)=\int dx_1\int dx_2  {\rm Tr}\left[i\omega^{\alpha}_{\ \beta}
M^{\beta}_F(x_1,x_2)
\left.\frac{\partial S^{\rm HP/SGP}_{\sigma,\mu\nu}(k_1,k_2,q,P_h/z)p^\sigma}
{\partial k^{\alpha}_2}\right|_{k_i=x_ip}\right], 
\label{polewmunu}
\eeq
where $\omega^\alpha_{\ \beta}=g^\alpha_{\ \beta}-p^\alpha n_\beta$ and 
the summation over color indices is implicit.  
$S^{\rm HP/SGP}_{\sigma,\mu\nu}$ is the partonic hard part
corresponding to the original nucleon matrix element
$M^\sigma (k_1,k_2)\sim{\cal FT}\la p|\bar{\psi} gA^\sigma \psi|p\ra$.  
Although we start from 
$\int d^4 k_1\int d^4k_2\,M^\sigma (k_1,k_2)S_\sigma (k_1,k_2)$ for the cross section,
reorganization of the collinear expansion using the Ward identity allows us to
convert the gauge field into the field strength in the correlation function and 
leads to the gauge invariant expression for the twist-3 cross section
as shown in (\ref{polewmunu}) (see \cite{Eguchi:2006mc} for the details).

The LO diagrams for 
$S^{\rm HP}_{\sigma,\mu\nu}$ are given in Fig. 2 of \cite{Eguchi:2006mc}
and the left two diagrams of Fig. 2 of \cite{Koike:2009yb}\footnote{Since both $E_F$ and $H_1$
are chiral-odd, right two diagrams in the same figure vanish. }.  
The LO diagrams for  $S^{\rm SGP}_{\sigma,\mu\nu}$ are given in Fig. 8 of \cite{Eguchi:2006mc}.  
We remind that the hard part for the HP contribution satisfies the following relation owing to
the Ward identity\cite{Eguchi:2006mc}, 
\beq
\left.
\frac{\partial S^{\rm HP}_{\sigma,\mu\nu}(k_1,k_2,q,P_h/z)p^\sigma}{\partial k_2^{\alpha}}\right|_{k_i=x_ip}
=\frac{1}{x_1-x_2}S^{\rm HP}_{\alpha,\mu\nu}(x_1p,x_2p,q,P_h/z),
\label{wardi}
\eeq
and thus one can obtain the HP contribution without calculating the derivative in (\ref{polewmunu}).  
The hard part contains the $\delta$-function corresponding to the on-shell 
condition for the final unobserved parton, which takes
\beq
\delta((xp+q-P_h/z)^2)=\frac{1}{Q^2\hat{z}}
\delta\left(\frac{q_T^2}{Q^2}-\left(1-\frac{1}{\hat{x}}\right)\left(1-\frac{1}{\hat{z}}\right)\right), 
\eeq
where the variables $\hat{x}$ and $\hat{z}$ are defined as
\beq
\hat{x}=\frac{x_{bj}}{x},\qquad	\hat{z}=\frac{z_f}{z}.  
\eeq
Calculating the LO diagrams for the hard part, we have obtained 
the HP contribution as
\beq
 &&\hspace{-0.5cm}\frac{d^{6}\Delta\sigma^{\rm HP}}{dx_{bj}dQ^{2}dz_{f}dq^{2}_{T}d\phi d\chi}  =
\frac{\alpha ^{2}_{em}\alpha_{s}}{16\pi^{2}S^{2}_{ep}x^{2}_{bj}Q^{2}}
\left(\frac{-\pi M_{N}}{4}\right)
\int \frac{dz}{z}H_1(z)\nn\\
&&
\times \int \frac{dx}{x}\left[
\frac{2}{1-\hat{x}}E_{F}\left(x_{bj},x\right)\left(\frac{4}{Nq_T}-
 \frac{4NQ^2(\hat{x}-1)}{q_T^3\hat{x}}\right)\sinh^2\psi  
\sin\left\{ \Phi_S +2(\phi-\chi)\right\}\right.
\nonumber\\[5pt]
&&\left.+E_F(x_{bj},x_{bj}-x)\left({-1\over N}\right)
\left\{  {8\hat{x}\over \hat{z} q_T}
(1+\cosh^2\psi)\sin\Phi_S \right.\right.\nn\\
&&\left.\left.
+  {8(1-\hat{x})Q\over \hat{z} q_T^2} \sinh 2\psi\sin(\Phi_s+\phi-\chi)
+  {8(1-\hat{x})^2Q^2\over \hat{x}\hat{z} q_T^3} \sinh^2\psi\sin\left\{\Phi_S+2(\phi-\chi)\right\}
\right\}
\right]\nn\\
 &&\quad\quad\quad\quad\times\delta\left(
 \frac{q_{T}^{2}}{Q^{2}}-\left(1-\frac{1}{\hat{x}}\right)
 \left(1-\frac{1}{\hat{z}}\right)\right), 
\label{HPform}
\eeq
where $\alpha_s=g^2/(4\pi)$ is the strong coupling constant.  
The $E_{F}\left(x_{bj},x\right)$ contribution occurs from
the diagrams in Fig. 2 of \cite{Eguchi:2006mc}
and the $E_F(x_{bj},x_{bj}-x)$ contribution is from
the left two diagrams of Fig. 2 of \cite{Koike:2009yb}.  
We note that $k=3,\ 8$ and $k=4,\ 9$ terms can be, respectively, combined into
the single sin forms $\sim \sin\left( \Phi_s +\phi-\chi\right)$
and $\sim \sin\left\{ \Phi_s +2(\phi-\chi)\right\}$. 

For the LO calculation of the SGP contribution we found the method using the master formula
\cite{Koike:2006qv} is convenient.  The result reads
\beq
 &&\hspace{-7mm}
 \frac{d^{6}\Delta\sigma^{\rm SGP}}{dx_{bj}dQ^{2}dz_{f}dq^{2}_{T}d\phi d\chi}\nonumber\\
 &&
 =\frac{\alpha ^{2}_{em}\alpha_{S}}{16\pi^{2}S^{2}_{ep}x^{2}_{bj}Q^{2}}\left(\frac{-\pi M_{N}}{4}\right)\frac{q_{T}}
{Q^{2}}\int \frac{dz}{z}H_1(z)\int \frac{dx}{x} \left(\frac{-1}{2N}\right)\nonumber\\
 && 
 \times\left[
 -8(1+\cosh^2\psi) \sin\Phi_S 
 \left\{ \frac{2\hat{x}}{1-\hat{z}}\left( x{dE_F(x,x)\over dx}-E_F(x,x)\right)
\right.\right.\nn\\
&&\left.\left.
\qquad\qquad 
+\left( {(1+\zh)Q^2\over \zh q_T^2}+{2\xh\over 1-\xh}\right)E_F(x,x)
 \right\}\right.\nn\\
 &&
 \left.
 +8\sinh 2\psi\sin(\Phi_S+\phi-\chi)
 \left\{ { 2\xh Q\over (1-\zh)q_T}\left( x{d E_F(x,x)\over dx} -E_F(x,x)\right)
 \right.\right.\nn\\
 &&
 \left.\left.\qquad\qquad
  -{Q\over 2q_T}\left( {Q^2\over q_T^2} +1\right) E_F(x,x)
 \right\}\right.\nn\\
 &&
 \left.
 -8\sinh^2\psi\sin\left\{\Phi_S+2(\phi-\chi)\right\}
 \left\{ { 2\xh Q^2\over (1-\zh)q_T^2}\left( x{d E_F(x,x)\over dx} -E_F(x,x)\right)
 \right.\right.\nn\\
 &&
 \left.\left.\qquad\qquad
 -\left(
 {3Q^4\over \zh q_T^4} + {Q^2\over q_T^2}\right) E_F(x,x)\right\}
 \right]\delta\left(
 \frac{q_{T}^{2}}{Q^{2}}-\left(1-\frac{1}{\hat{x}}\right)
 \left(1-\frac{1}{\hat{z}}\right)\right).
\label{SGPform}
\eeq
It turned out that the LO SGP cross sections for $k=3,\ 8$ and $k=4,\ 9$
can be also transformed into the single $\sin$ forms as in the HP contribution. 

To summarize this subsection, the cross section in (\ref{EFH1contribution}) 
is given by the sum of (\ref{HPform})
and (\ref{SGPform}).

\subsection{Contribution from twist-3 quark fragmentation function}

\begin{figure}[h]
\begin{center}
\includegraphics[width=16cm]{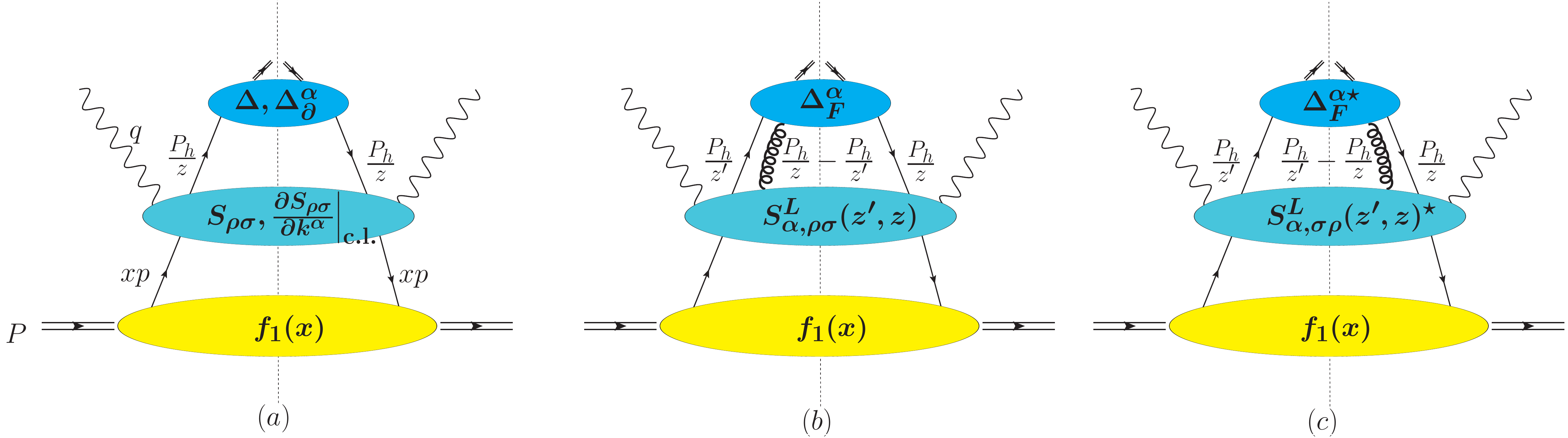}\hspace{1cm}
\end{center}
\caption{Twist-3 quark fragmentation function contribution to $ep\to e\Lambda^\uparrow X$.
The top blob represents the fragmentation correlator, the bottom one indicates
the unpolarized quark DF, and the middle one corresponds to the partonic hard part.
The contribution from the unpolarized gluon DF should also be included.  }
\label{SIDISqfragFig}
\end{figure}

The formalism for calculating the twist-3 quark FF contribution
has been developed in \cite{Kanazawa:2013uia}
for the process $ep^\uparrow \to e\pi X$.  
This contribution is diagrammatically shown in Fig. \ref{SIDISqfragFig}.  
From eq. (54) of \cite{Kanazawa:2013uia}, one 
can write the twist-3 quark FF contribution to (\ref{sidishyperon}) as
\begin{eqnarray}
&&\hspace{-0.8cm}
\frac{d^6\Delta\sigma^{\rm tw3-frag}}{dx_{bj}dQ^2dz_fdq_T^2d\phi d\chi}
=\frac{\alpha_{em}^{2}}{128\pi^4x_{bj}^2S_{ep}^2Q^2}z_fQ^2
\sum_{k=1,\cdots,4,8,9}\mathcal{A}_{\it k}(\phi-\chi)
\int\frac{dx}{x}f_1(x)\nn\\[5pt]
&&\  \times
\left[
\int\frac{dz}{z^2}{\rm Tr}\left[\Delta(z)
S_{\rho\sigma}(z)\right]
\widetilde{\mathscr{V}}_k^{\rho\sigma}
+\Omega^{\alpha}_{\ \beta}\int\frac{dz}{z^2}
{\rm Tr}
\left[\Delta^{\beta}_{\del}(z)\left.\frac{\del S_{\rho\sigma}(k)}{\del k^{\alpha}}\right|_{c.l.}\right]
\widetilde{\mathscr{V}}_k^{\rho\sigma}
\right.
\nn\\[5pt]
&&\left.\  +\Omega^{\alpha}_{\ \beta}\int\frac{dz}{z^2}\int\frac{dz'}{{z'}^2}
{\rm P}\left(\frac{1}{1/z'-1/z}\right)
\left\{
\Im
{\rm Tr}\left[\Delta^{\beta}_F(z,z')S^{L}_{\alpha, \rho\sigma}(z',z)\right]+(\rho\leftrightarrow\sigma)
\right\}
\widetilde{\mathscr{V}}_k^{\rho\sigma}
\right],
\label{SIDISqFraForma}
\end{eqnarray}
where $\Omega^\alpha_{\ \beta} =g^\alpha_{\ \beta} -P_h^\alpha w_\beta$, 
$f_1(x)$ denotes the twist-2 unpolarized quark or gluon distribution function, and
the summation over quark flavors as well as the factor associated with
the quark's fractional electric charge is omitted.  
In (\ref{SIDISqFraForma}), $S_{\rho\sigma}(k)$ and $S^{L}_{\alpha, \rho\sigma}(z',z)$ 
are the partonic hard parts originally associated,
respectively, with the fragmentation matrix elements
$\sum_X\la 0| \psi|hX\ra\la hX|\bar{\psi}|0\ra$ and
$\sum_X\la 0| \psi|hX\ra\la hX|\bar{\psi}gA^\alpha|0\ra$.  
Inserting (\ref{qFraI}), (\ref{qFraK}) and (\ref{qFraD}) into (\ref{SIDISqFraForma}), 
one can obtain the twist-3 FF contribution to the cross section 
as\footnote{The $T$-even functions 
$G_T(z)$ in (\ref{qFraI}) and $G_{1T}^{(1)\perp} (z)$ in (\ref{qFraK}) 
do not contribute to the symmetric part of the hadronic tensor.}
\begin{eqnarray}
&&\hspace{-0.7cm}
\frac{d^6\Delta\sigma^{\rm tw3-frag}}{dx_{bj}dQ^2dz_fdq_T^2d\phi d\chi}\nonumber\\[5pt]
&&=\frac{\alpha_{em}^{2}\alpha_s(-M_h)}{16\pi^2x_{bj}^2S_{ep}^2Q^2}
\sum_{k=1}^9
\mathcal{A}_{\it k}(\phi-\chi)\mathcal{S}_{\it k}\int\frac{dx}{x}f_1(x)\int\frac{dz}{z}\delta\left(\frac{q_T^2}{Q^2}-\left(1-\frac{1}{\xh}\right)\left(1-\frac{1}{\zh}\right)\right)\nn\\[5pt]
&&\times\Biggl[
\frac{D_T(z)}{z}\hat{\sigma}^{k}_{int}-\left\{\frac{d}{d(1/z)}\frac{D^{\perp(1)}_{1T}(z)}{z}\right\}\hat{\sigma}^{k}_{kin1}-D^{\perp(1)}_{1T}(z)\hat\sigma^{k}_{kin2}\nn\\[5pt]
&&+\int\frac{dz'}{{z'}^2}{\rm P}\left(\frac{1}{1/z-1/z'}\right)\biggl\{
{\Im}\widehat{D}_{FT}(z,z')\Bigl[\hat{\sigma}_{DF3}^{k}-\frac{2}{z}
\left(\frac{1}{1/z-1/z'}\right)
\hat{\sigma}_{DF4}^{k}\nn\\[5pt]
&&\qquad\qquad\qquad\qquad\qquad
+\frac{z'}{z}\hat{\sigma}_{DF1}^{\it k}+\frac{1}{z}\left(\frac{1}{z'}-\frac{1}{\left(1-q_T^2/Q^2\right)z_f}\right)^{-1}
\hat{\sigma}_{DF2}^{k}\Bigr]\nn\\[5pt]
&&\qquad\qquad
+{\Im}\widehat{G}_{FT}(z,z')\Bigl[\hat{\sigma}_{GF3}^{\it k}-\frac{2}{z}\left(\frac{1}{1/{z}-1/z'}\right)
\hat{\sigma}_{GF4}^{\it k}+\frac{z'}{z}\hat{\sigma}_{GF1}^{\it k}\nn\\[5pt]
&&\qquad\qquad\qquad\qquad\qquad
+\frac{1}{z}\left(\frac{1}{z'}-\frac{1}{\left(1-q_T^2/Q^2\right)z_f}\right)^{-1}
\hat{\sigma}_{GF2}^{\it k}\Bigr]
\biggr\}\Biggr],
\label{SIDISqfragForma2}
\end{eqnarray}
where $\mathcal{S}_{1,2,3,4}=\sin\Phi_S$ and $\mathcal{S}_{8,9}=\cos\Phi_S$, and 
$\hat{\sigma}^{k}_{int}$, $\hat\sigma^{k}_{kin1}$, $\cdots$ etc. represent the
partonic hard cross sections. 
In (\ref{SIDISqfragForma2}) we have explicitly separated the $z'$-dependence 
of the cross section for the dynamical FFs, 
and introduced the $z'$-independent partonic hard cross section, 
$\hat{\sigma}^k_{DF1}$, $\hat{\sigma}^k_{DF2}$, $\hat{\sigma}^k_{DF3}$, $\hat{\sigma}^k_{DF4}$
for $\Im \widehat{D}_{FT}(z,z')$, and likewise for $\Im \widehat{G}_{FT}(z,z')$.  

From actual calculation, it is easy to find
\beq
\hat{\sigma}^k_{DF3}=-\hat{\sigma}^k_{GF3},\qquad\hat{\sigma}^k_{GF4}=0. 
\label{Xsecrel1}
\eeq
Thanks to these relations, 
the EOM relation (\ref{qFra_EOM}) and the LIR (\ref{qFra_LIR}) allow one to
absorb the contribution from the dynamical FFs with,
$\hat{\sigma}^k_{DF3}$,
$\hat{\sigma}^k_{DF4}$,
$\hat{\sigma}^k_{GF3}$ and
$\hat{\sigma}^k_{GF4}$, 
into those from the intrinsic and the kinematical FFs.
Defining the new partonic hard cross sections, 
\begin{eqnarray}
&&\hat{\sigma}^k_T\equiv\hat{\sigma}_{int}^k+\hat{\sigma}_{DF3}^k+
\hat{\sigma}_{DF4}^k,\\
&&\hat{\sigma}_{\perp D}^{k}\equiv\hat{\sigma}_{kin1}^{k}-\hat{\sigma}_{DF4}^k,\\
&&\hat{\sigma}_{\perp}^{k}\equiv\hat{\sigma}_{kin2}^{k}-\hat{\sigma}_{DF3}^k, 
\label{defXsec1}
\end{eqnarray}
one can rewrite (\ref{SIDISqfragForma2}) in the following form: 
\begin{eqnarray}
&&\hspace{-0.7cm}
\frac{d^6\Delta\sigma^{\rm tw3-frag}}{dx_{bj}dQ^2dz_fdq_T^2d\phi d\chi}\nn\\
&&=\frac{\alpha_{em}^{2}\alpha_s(-M_h)}{16\pi^2x_{bj}^2S_{ep}^2Q^2}
\sum_{k=1}^9
\mathcal{A}_{k}(\phi-\chi)\mathcal{S}_{k}\int\frac{dx}{x}f_1(x)
\int\frac{dz}{z}\delta\left(\frac{q_T^2}{Q^2}-\left(1-\frac{1}{\xh}\right)
\left(1-\frac{1}{\zh}\right)\right)\nn\\
&&\times\Biggl[
\frac{D_T(z)}{z}\hat{\sigma}^{k}_T-\left\{\frac{d}{d(1/z)}
\frac{D^{\perp(1)}_{1T}(z)}{z}\right\}
\hat{\sigma}^{\it k}_{\perp D} -D^{\perp(1)}_{1T}(z)\hat\sigma^{\it k}_\perp\nn\\
&&+\int\frac{dz'}{{z'}^2}{\rm P}\left(\frac{1}{1/z-1/z'}\right)\Biggl\{{\Im}\widehat{D}_{FT}(z,z')
\left[\frac{z'}{z}\hat{\sigma}_{DF1}^{k}+
\frac{1}{z}\left(\frac{1}{z'}-
\frac{1}{\left(1-q_T^2/Q^2\right)z_f}\right)^{-1}\hspace{-3mm}\hat{\sigma}_{DF2}^{k}\right]\nn\\
&&+{\Im}\widehat{G}_{FT}(z,z')\left[
\frac{z'}{z}\hat{\sigma}_{GF1}^{k}+\frac{1}{z}
\left(\frac{1}{z'}-\frac{1}{\left(1-q_T^2/Q^2\right)z_f}\right)^{-1}\hspace{-3mm}
\hat{\sigma}_{GF2}^{\it k}\right]\Biggr\}
\Biggr].
\label{qFraXsecfinal}
\end{eqnarray}
This is the final form we use to present the twist-3 quark FF contribution to 
$ep\to e\Lambda^\uparrow X$ shown in Fig. \ref{SIDISqfragFig}.  

In order to present the partonic hard cross sections in (\ref{qFraXsecfinal}), 
it is convenient to introduce
the quark FF contribution to the twist-2 unpolarized cross section 
for $ep\to e\Lambda X$.  It reads
\begin{eqnarray}
&&\frac{d^6\sigma^{\rm unpol}}{dx_{bj}dQ^2dz_fdq_T^2d\phi d\chi}
=\frac{\alpha_{em}^{2}\alpha_s}{16\pi^2x_{bj}^2S_{ep}^2Q^2}\sum_{k=1}^4
\mathcal{A}_k(\phi-\chi)\nn\\[5pt]
&&\qquad\times
\int\frac{dx}{x}f_1(x)\int\frac{dz}{z}D_q(z)\delta\left(\frac{q_T^2}{Q^2}
-\left(1-\frac{1}{\xh}\right)\left(1-\frac{1}{\zh}\right)\right)  C_F\hat{\sigma}_{U}^k, 
\label{twist2Xsec}
\end{eqnarray}
where $D_q(z)$ denotes the twist-2 unpolarized quark FF, and
$C_F\hat{\sigma}_{U}^k$ represents the partonic hard cross sections
which depend on whether $f_1(x)=q(x)$ (quark DF) or $f_1(x)=G(x)$ (gluon DF).  
We will see below that some of the hard cross sections in 
(\ref{qFraXsecfinal}) are related to $\hat{\sigma}_{U}^k$s.

Below we give LO Feynman diagrams for the hard part and the results for the hard cross sections
$\hat{\sigma}_{T,\perp D, \perp}$ and 
$\hat{\sigma}_{DF1,GF1,DF2,GF2}$, which are the functions of
$Q$, $q_T$, $\xh=x_{bj}/x$ and $\zh=z_f/z$.  

\vspace{1cm}

\begin{figure}[h]
\begin{center}
\includegraphics[width=8.0cm]{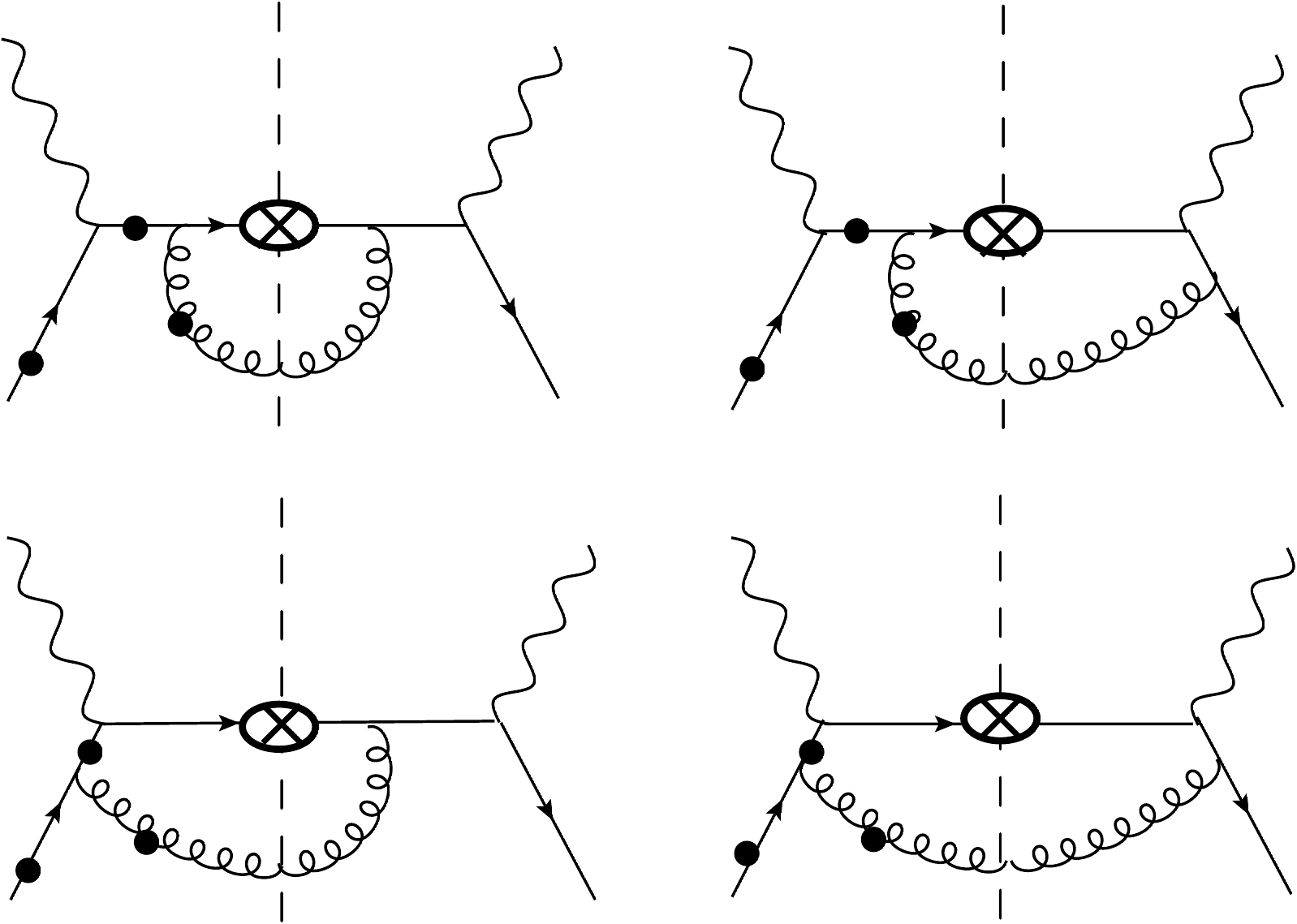}
\end{center}
\caption{LO diagrams in the $\gamma q\to qg$ channel contributing to the hard parts
$S_{\rho\sigma}(k)$ and 
$S^{L}_{\alpha,\rho\sigma}(z',z)$ in (\ref{SIDISqFraForma}).  A cross in an oval in each diagram
represents the fragmentation insertion. 
For $S^{L}_{\alpha,\rho\sigma}(z',z)$, an extra gluon line connecting the cross in the oval
and each dot in the diagram should be added.  }
\label{gqqg_diagram}
\end{figure}

\noindent
(I) $\gamma q\to qg$ channel:

The diagrams for the hard part in this channel are shown in Fig. \ref{gqqg_diagram}.  
It is convenient to present some of
the hard cross sections in (\ref{qFraXsecfinal}) in the following form:  
\begin{eqnarray}
&&\hat{\sigma}_{T}^{k}=C_F\hat{\sigma}_{1}^{k}-\frac{1}{N}\frac{1-\hz}{\qt\hz}
\hat{\sigma}_{U}^{\it k},
\label{sigmaTq}\\
&&\hat{\sigma}_{\perp D}^{k}
=\frac{1}{N}\frac{1-\hz}{\qt\hz}\hat{\sigma}_{U}^{k},
\label{sigmaperpDq}\\
&&\hat{\sigma}_{\perp}^{k}=C_F\hat{\sigma}_{2}^{k},\\
&&\hat{\sigma}_{DF2}^{k}
=(1-\hz)\left(\frac{C_F}{1-\hx-\hz}-\frac{1}{2 N \hz}\right)\hat{\sigma}_{2}^{k},\\
&&\hat{\sigma}_{GF2}^{k}
=(1-\hz)\left(\frac{C_F}{1-\hx-\hz}-\frac{1}{2 N \hz}\right)\hat{\sigma}_{3}^{k}, 
\end{eqnarray}
where $\hat\sigma^k_U$s are the unpolarized partonic cross sections introduced in
(\ref{twist2Xsec}) for the unpolarized quark distribution $f_1(x)=q(x)$ and 
$\hat{\sigma}^k_{1,2,3}$ are newly introduced partonic cross sections, 
and hence specification of 
$\hat{\sigma}_{U}^{k}$, 
$\hat{\sigma}^k_{1,2,3}$, $\hat{\sigma}_{DF1}$ and $\hat{\sigma}_{GF1}$
determines the cross sections in (\ref{qFraXsecfinal}).  
We shall now give those in the $\gamma q\to qg$ channel. 

\vspace{0.5cm}

\vspace{0.3cm}

\noindent
(1) Twist-2 unpolarized cross section:
\begin{eqnarray}
&&\hat{\sigma}_{U}^{\it 1}=
\frac{2 (2 - 2 \hx + \hx^2 + \hz (-2 + 8 \hx - 6 \hx^2) 
+ \hz^2 (1 - 6 \hx + 6 \hx^2))}{(1 - \hz) (1 - \hx)},\nn\\[5pt]
&&\hat{\sigma}_{U}^{\it 2}=8 \hx \hz,\nn\\[5pt]
&&\hat{\sigma}_{U}^{\it 3}=\frac{4 Q (1 - \hx + \hz (-1 + 2 \hx))}{\qt},\nn\\[5pt]
&&\hat{\sigma}_{U}^{\it 4}=4 \hx \hz.
\end{eqnarray}
$\hat{\sigma}_{U}^{\it 8,9}=0$ is understood in (\ref{sigmaTq}) and (\ref{sigmaperpDq}).  

\vspace{0.5cm}

\noindent
(2) $k={\it 1}$:
\begin{eqnarray}
&&\hspace{-0.7cm}\hat\sigma^{\it 1}_{1}=\frac{4 (1 - 9 \hx + 14 \hx^2 - 6 \hx^3 
+ 2 \hz^2 (1 - 6 \hx + 6 \hx^2) + 
\hz (-4 + 23 \hx - 30 \hx^2 + 12 \hx^3)) }{\qt (-1 + \hx) (-1 + \hz + \hx)},\nn\\[5pt]
&&\hspace{-0.7cm}\hat\sigma^{\it 1}_{2}=-\frac{4 ((-1 + \hx)^3 - 3 \hz (-1 + \hx)^2 (-1 + 2 \hx) 
+ \hz^3 (1 - 6 \hx + 6 \hx^2) + 
\hz^2 (-3 + 16 \hx - 18 \hx^2 + 6 \hx^3))}{\qt (-1 + \hz) (-1 + \hx) (-1 + \hz + \hx)},\nn\\[5pt]
&&\hspace{-0.7cm}\hat\sigma^{\it1}_{3}
=-\frac{4 ((-1 + \hx)^3 - 3 \hz (-1 + \hx)^2 (-1 + 2 \hx) + 
\hz^3 (1 - 6 \hx + 6 \hx^2) + 
\hz^2 (-3 + 14 \hx - 18 \hx^2 + 6 \hx^3))}{\qt (-1 + \hx) (-1+\hz) (-1 + \hz + \hx)},\nn\\[5pt]
&&\hspace{-0.7cm}\hat\sigma^{\it 1}_{DF1}=
\frac{1}{N}\frac{2 (-1 + (-1 + \hz) \hx)}{\qt \hz (-1 + \hx)},\nn\\[5pt]
&&\hspace{-0.7cm}\hat\sigma^{\it 1}_{GF1}=
-\frac{1}{N}\frac{2 (1 + (-1 + \hz) \hx)}{\qt \hz (-1 + \hx)}, 
\end{eqnarray}

\vspace{0.5cm}

\noindent
(3) $k={\it 2}$: 
\begin{eqnarray}
&&\hat\sigma^{{\it 2}}_{1}=\frac{16 (-1 + 2 \hz) \hx}{\qt},\nn\\[5pt]
&&\hat\sigma^{{\it 2}}_{2}=-\frac{16 \hx \hz}{\qt},\nn\\[5pt]
&&\hat\sigma^{{\it 2}}_{3}=-\frac{16 \hz \hx}{\qt},\nn\\[5pt]
&&\hat\sigma^{\it 2}_{DF1}=0,\nn\\[5pt]
&&\hat\sigma^{\it 2}_{GF1}=0.
\end{eqnarray}

\vspace{0.5cm}

\noindent
(4) $k={\it 3}$:
\begin{eqnarray}
&&\hat\sigma^{{\it 3}}_{1}=
\frac{4 \hx ((-1 + \hx)^2 + \hz^3 (-4 + 8 \hx) + 2 \hz^2 (5 - 10 \hx + 4 \hx^2) - 
   2 \hz (3 - 7 \hx + 4 \hx^2))}{Q(-1 + \hz)(-1 + \hx) (-1 + \hz + \hx)},\nn\\[5pt]
&&\hat\sigma^{{\it 3}}_{2}=
-\frac{4 \hx \hz (-2 (-1 + \hx)^2 + \hz^2 (-2 + 4 \hx) + 
\hz (5 - 8 \hx + 4 \hx^2)) }{Q(-1 + \hz) (-1 + \hx) (-1 + \hz + \hx)},\nn\\[5pt]
&&\hat\sigma^{{\it 3}}_{3}
=-\frac{4 Q (-2 (-1 + \hx)^2 + \hz^2 (-2 + 4 \hx) + \hz (3 - 8 \hx + 4 \hx^2)) }{\qt^2 (-1 + \hz + \hx)},\nn\\[5pt]
&&\hat\sigma^{\it 3}_{DF1}=\frac{1}{N}\frac{2 \hx}{Q (-1+\hx)},\nn\\[5pt]
&&\hat\sigma^{\it 3}_{GF1}=-\frac{1}{N}\frac{2 \hx}{Q (-1+\hx)}.
\end{eqnarray}

\vspace{0.5cm}

\noindent
(5) $k={\it 4}$:
\begin{eqnarray}
&&\hat\sigma^{{\it 4}}_{1}=4 \frac{ (4 \hx \hz^2 - 2 (-1 + \hx) x + \hz (1 - 6 \hx + 4 \hx^2))}{\qt (-1 + \hz + \hx)},\nn\\[5pt]
&&\hat\sigma^{{\it 4}}_{2}=-\frac{4 \hz (1 + 2 (-1 + \hz) \hx + 2 \hx^2)}{\qt (-1 + \hz + \hx)},\nn\\[5pt]
&&\hat\sigma^{{\it 4}}_{3}=-\frac{4 \hz (-1 + 2 (-1 + \hz) \hx + 2 \hx^2)}{\qt (-1 + \hz + \hx)},\nn\\[5pt]
&&\hat\sigma^{\it 4}_{DF1}=\frac{1}{N}\frac{2}{\qt},\nn\\[5pt]
&&\hat\sigma^{\it 4}_{GF1}=-\frac{1}{N}\frac{2}{\qt}.
\end{eqnarray}

\vspace{0.5cm}

\noindent
(6) $k={\it 8}$:
\begin{eqnarray}
&&\hat\sigma^{{\it 8}}_{1}=\frac{4 \hx (-(-1 + \hx)^2 + 2 \hz (-1 + \hx)^2 + 2 \hx \hz^2)}{Q(-1 + \hz) (-1 + \hx) (-1 + \hz + \hx)},\nn\\[5pt]
&&\hat\sigma^{{\it 8}}_{2}=-\frac{4 \hx \hz^2}{Q(-1 + \hz) (-1 + \hx) (-1 + \hz + \hx)},\nn\\[5pt]
&&\hat\sigma^{{\it 8}}_{3}=\frac{4 \hz Q}{\qt^2 (-1 + \hz + \hx)},\nn\\[5pt]
&&\hat\sigma^{\it 8}_{DF1}=-\frac{1}{N}\frac{2 \hx}{Q (1-\hx)},\nn\\[5pt]
&&\hat\sigma^{\it 8}_{GF1}=\frac{1}{N}\frac{2 \hx}{Q (1-\hx)}. 
\end{eqnarray}

\vspace{0.5cm}

\noindent
(7) $k={\it 9}$:
\begin{eqnarray}
&&\sigma^{{\it 9}}_{1}=\frac{4 \left\{\hz + 2 \hx \hz + 2 (-1 + \hx) \hx\right\}}{\qt (-1 + \hz + \hx)},\nn\\[5pt]
&&\sigma^{{\it 9}}_{2}=-\frac{4 \hz}{\qt (-1 + \hz + \hx)},\nn\\[5pt]
&&\sigma^{{\it 9}}_{3}=\frac{4 \hz}{\qt (-1 + \hz + \hx)},\nn\\[5pt]
&&\sigma^{\it 9}_{DF1}=\frac{1}{N}\frac{2}{\qt},\nn\\[5pt]
&&\sigma^{\it 9}_{GF1}=-\frac{1}{N}\frac{2}{\qt}.
\end{eqnarray}

\vspace{1cm}

\begin{figure}[h]
  \begin{center}
   \includegraphics[width=10.0cm]{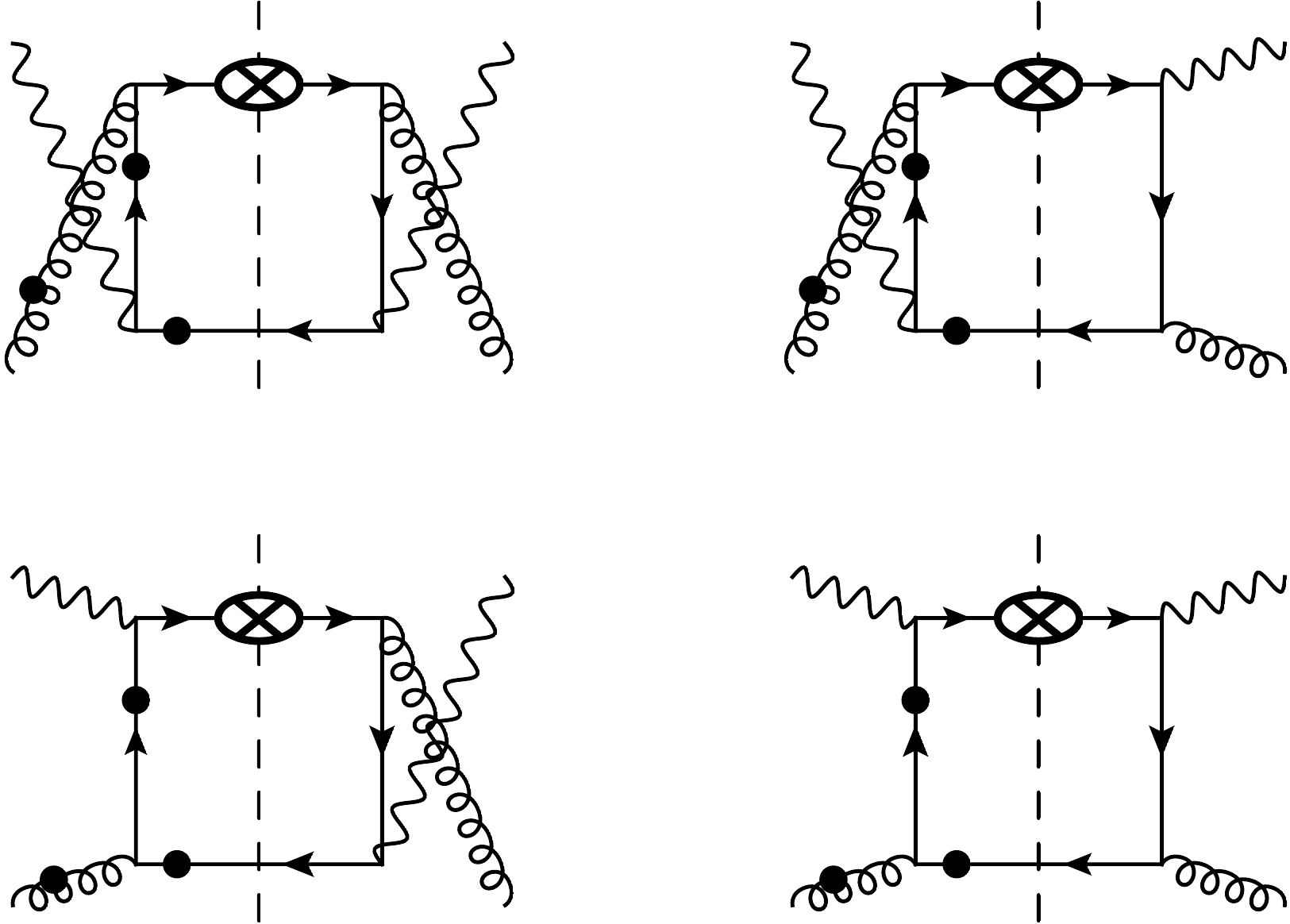}
  \end{center}
\caption{
LO diagrams 
in the $\gamma g\to q\bar{q}$ channel contributing to the hard parts
$S_{\rho\sigma}(k)$ and 
$S^{L}_{\alpha,\rho\sigma}(z',z)$ in (\ref{SIDISqFraForma}).  A cross in an oval in each diagram
represents the fragmentation insertion. 
For $S^{L}_{\alpha,\rho\sigma}(z',z)$, an extra gluon line connecting the cross in the oval
and each dot in the diagram should be added.  
\label{ggqq_diagram}
}
\end{figure}

\noindent
(II) $\gamma g\to q\bar{q}$ channel:


Diagrams for the hard part in this channel are shown in Fig. \ref{ggqq_diagram}.  
It is convenient to present some of the hard cross sections in 
(\ref{qFraXsecfinal}) in the following form:  
\begin{eqnarray}
&&\hat{\sigma}_{T}^{\it k}=\hat{\sigma}_{1}^{{\it k}}-\frac{1}{N^2-1}\frac{1-\hz}{\qt\hz}\hat{\sigma}_{U}^{\it k},\label{sigmaTg}\\[5pt]
&&\hat{\sigma}_{\perp D}^{\it k}=
\frac{N}{2C_F}\frac{1-\hz}{\qt\hz}\hat{\sigma}_{U}^{\it k},\label{sigmaperpDg}\\[5pt]
&&\hat{\sigma}_{DF2}^{\it k}=
\frac{2\hx}{1-\hx-\hz}\left(\frac{1-\hx}{1-\hx-\hz}+
\frac{1}{N^2-1}\right)\hat{\sigma}_{2}^{{\it k}},\\[5pt]
&&\hat{\sigma}_{GF2}^{\it k}=
\frac{2\hx}{1-\hx-\hz}\left(\frac{1-\hx}{1-\hx-\hz}+
\frac{1}{N^2-1}\right)\hat{\sigma}_{3}^{{\it k}},
\end{eqnarray}
where $\hat\sigma^k_U$'s are the unpolarized partonic cross sections in (\ref{twist2Xsec})
for the unpolarized gluon distribution $f_1(x)=G(x)$,
and $\hat\sigma^k_{1,2,3}$ are newly introduced here.
We will give 
$\hat\sigma^k_U$, $\hat\sigma^k_{1,2,3}$, $\hat\sigma^k_\perp$, $\hat\sigma^k_{DF1}$
and $\hat\sigma^k_{GF1}$ below.  

\vspace{0.5cm}

\vspace{0.3cm}

\noindent
(1) Twist-2 unpolarized cross sections:
\begin{eqnarray}
&&\hat{\sigma}_{U}^{\it 1}=2\frac{1 - 2 \hx + 2 \hx^2 - 2 \hz (1 - 6 \hx + 6 \hx^2) + 2 \hz^2 (1 - 6 \hx + 6 \hx^2)}{\hz(1 - \hz)},\nn\\[5pt]
&&\hat{\sigma}_{U}^{\it 2}=16 \hx (1-\hx),\nn\\[5pt]
&&\hat{\sigma}_{U}^{\it 3}=\frac{8 \hx\qt(1 - 2 \hz) (1 - 2 \hx)}{Q(1 - \hz)},\nn\\[5pt]
&&\hat{\sigma}_{U}^{\it 4}=8 \hx (1-\hx),
\end{eqnarray}
where $\hat{\sigma}_{U}^{\it 8,9}=0$ is understood in (\ref{sigmaTg}) and (\ref{sigmaperpDg}).

\vspace{0.5cm}
 
 \noindent
(2) $k=1$:
\begin{eqnarray}
&&\hat\sigma^{{\it 1}}_{1}=-\frac{2 (-1 +\hx) \{-2 + 5 \hx - 3 \hx^2 + \hz^2 (-1 + 6 \hx^2) + 
   \hz (4 - 11 \hx + 6 \hx^2)\} }{\qt \hz^2 (-1 + \hz + \hx)},\nn\\[5pt]
&&\hat\sigma^{\it 1}_{\perp}=\frac{2}{\qt (-1 + \hz) \hz^2 (-1 + \hz + \hx)} 
\left\{\hx - \hx^3 + 2 \hz^4 (1 - 6 \hx + 6 \hx^2) + 
    \hz^2 (4 - 30 \hx + 55 \hx^2 - 24 \hx^3) \right.\nn\\[5pt]
   &&\left. \qquad\qquad\qquad+  \hz (-1 + 5 \hx - 15 \hx^2 + 9 \hx^3) + 
     \hz^3 (-5 + 37 \hx - 54 \hx^2 + 18 \hx^3)\right\},\nn\\[10pt]
&&\hat\sigma^{\it 1}_{DF1}=-\frac{1}{N^2-1}\frac{2(3 \hz (1 - 2 \hx) \hx + \hx (1 + \hx) + \hz^2 (1 - 6 \hx + 6 \hx^2))}{\qt \hz^2},\nn\\[5pt]
&&\hat\sigma^{\it 1}_{GF1}=-\frac{1}{N^2-1}\frac{2(\hz (5 - 6 \hx) \hx + (-1 + \hx) \hx + \hz^2 (1 - 6 \hx + 6 \hx^2)) }{\qt \hz^2},\nn\\[5pt]
&&\hat\sigma^{{\it 1}}_{2}=\frac{1}{q_T\hx\hz^2}
\left\{
(-1 + \hx)^2 (-1 + 2 \hx) - 3 \hz (-1 + \hx)^2 (-1 + 4 \hx) \right.\nn\\
&&\left.\qquad\qquad\qquad\qquad
+\hz^3 (1 - 6 \hx + 6 \hx^2) + 3 \hz^2 (-1 + 7 \hx - 10 \hx^2 + 4 \hx^3)\right\},\nn\\[10pt]
&&\hat\sigma^{{\it 1}}_{3}=\frac{-(-1 + \hx)^2 + 3 \hz (-1 + \hx)^2 + 
   \hz^2 (-3 + 9 \hx - 6 \hx^2) + \hz^3 (1 - 6 \hx + 6 \hx^2)}{q_T\hx\hz^2}.
\end{eqnarray}

\vspace{0.5cm}

\noindent
(3) $k=2$:
\begin{eqnarray}
&&\hat\sigma^{{\it 2}}_{1}=
-\frac{8 (-1 + \hx) \hx (-1 + \hx + \hz \hx) }{\qt \hz (-1 + \hz + \hx)},\nn\\[5pt]
&&\hat\sigma^{\it 2}_{\perp}
=\frac{8 (-1 + \hx) \hx (1 + 2 \hz^2 - \hx + \hz (-4 + 3 \hx))}{\qt 
\hz (-1 + \hz + \hx)},\nn\\[5pt]
&&\hat\sigma^{\it 2}_{DF1}=-\frac{1}{N^2-1}\frac{8 \qt \hx^2}{Q^2},\nn\\[5pt]
&&\hat\sigma^{\it 2}_{GF1}=-\frac{1}{N^2-1}\frac{8 \qt \hx^2}{Q^2},\nn\\[5pt]
&&\hat\sigma^{{\it 2}}_{2}={4 \qt \hx (-2 + \hz + 2 \hx)\over Q^2},\nn\\[5pt]
&&\hat\sigma^{{\it 2}}_{3}={4 \qt \hx \hz\over Q^2}.
\end{eqnarray}

\vspace{0.5cm}

\noindent
(4) $k=3$:
\begin{eqnarray}
&&\hat\sigma^{{\it 3}}_{1}
=-\frac{2 \hx (-4 (-1 + \hx)^2 + 2 \hz (3 - 4 \hx + \hx^2) + \hz^2 (-1 - 2 \hx + 4 \hx^2))}{Q(-1 + \hz) \hz (-1 + \hz + \hx)},\nn\\[5pt]
&&\hat\sigma^{\it 3}_{\perp}
=\frac{2 \hx (-1 - \hx + 2 \hx^2 + \hz^3 (-4 + 8 \hx) + \hz (-3 + 16 \hx - 10 \hx^2) + 
   \hz^2 (9 - 26 \hx + 12 \hx^2))}{Q(-1 + \hz) \hz (-1 + \hz + \hx)},\nn\\[5pt]
&&\hat\sigma^{\it 3}_{DF1}=-\frac{1}{N^2-1}\frac{2\hx (-1 - 2 \hx + \hz (-2 + 4 \hx))}{Q\hz},\nn\\[5pt]
&&\hat\sigma^{\it 3}_{GF1}=-\frac{1}{N^2-1}\frac{2(-1 + 2 \hz) \hx (-1 + 2 \hx)}{Q\hz},\nn\\[5pt]
&&\hat\sigma^{{\it 3}}_{2}=\frac{ -3 + 7 \hx - 4 \hx^2 + \hz^2 (-2 + 4 \hx) + 2 \hz (3 - 7 \hx + 4 \hx^2)}{Q\hz},\nn\\[5pt]
&&\hat\sigma^{{\it 3}}_{3}=\frac{ -1 - 2 \hz (-1 + \hx) + \hx + \hz^2 (-2 + 4 \hx)}{Q\hz}.
\end{eqnarray}

\vspace{0.5cm}

\noindent
(5) $k=4$:
\begin{eqnarray}
&&\hat\sigma^{{\it 4}}_{1}=
-\frac{2 (-1 + \hx) (1 - 3 \hx + 2 (1 + \hz) \hx^2)}{\qt \hz (-1 + \hz + \hx)},\nn\\[5pt]
&&\hat\sigma^{\it 4}_{\perp}=\frac{2 (-1 + \hx) (1 + (1 - 8 \hz + 4 \hz^2) x + (-2 + 6 \hz) \hx^2)}{\qt \hz (-1 + \hz + \hx)},\nn\\[5pt]
&&\hat\sigma^{\it 4}_{DF1}=-\frac{1}{N^2-1}\frac{2(-1 + \hx) (-1 + 2 (-1 + \hz) \hx)}{\qt \hz},\nn\\[5pt]
&&\hat\sigma^{\it 4}_{GF1}=-\frac{1}{N^2-1}\frac{2(-1 + \hx) (1 + 2 (-1 + \hz) \hx)}{\qt\hz},\nn\\[5pt]
&&\hat\sigma^{{\it 4}}_{2}={\qt (1 + 2 (-2 + \hz) \hx + 4 \hx^2)\over Q^2},\nn\\[5pt]
&&\hat\sigma^{{\it 4}}_{3}={\qt (-1 + 2 \hx \hz)\over Q^2}.
\end{eqnarray}

\vspace{0.5cm}

\noindent
(6) $k=8$:
\begin{eqnarray}
&&\hat\sigma^{{\it 8}}_{1}=-\frac{2 \hx (-2 (-1 + \hx)^2 + 4 \hz (-1 + \hx)^2 + \hz^2 (-1 + 2 \hx))}{Q(-1 + \hz) \hz (-1 + \hz + \hx)},\nn\\[5pt]
&&\hat\sigma^{\it 8}_{\perp}=-\frac{2 \hx (1 - \hz - \hx + \hz^2 (-1 + 2 \hx))}{Q(-1 + \hz) \hz (-1 + \hz + \hx)},\nn\\[5pt]
&&\hat\sigma^{\it 8}_{DF1}=\frac{1}{N^2-1}\frac{\hx}{2Q\hz },\nn\\[5pt]
&&\hat\sigma^{\it 8}_{GF1}=-\frac{1}{N^2-1}\frac{\hx}{2Q\hz },\nn\\[5pt]
&&\hat\sigma^{{\it 8}}_{2}=-\frac{(-1 + 2 \hz)  (-1 + \hx)^2}{Q\hz},\nn\\[5pt]
&&\hat\sigma^{{\it 8}}_{3}=\frac{(-1 + 2 \hz)  (-1 + \hx)}{Q\hz}.
\end{eqnarray}

\vspace{0.5cm}

\noindent
(7) $k=9$:
\begin{eqnarray}
&&\hat\sigma^{{\it 9}}_{1}=-\frac{2 (-1 + \hx) (1 + (-5 + 2 \hz) \hx + 4 \hx^2)}{\qt \hz (-1 + \hz + \hx)},\nn\\[5pt]
&&\hat\sigma^{\it 9}_{\perp}=-\frac{2 (-1 + \hx) (-1 + \hx + 2 \hz \hx)}{\qt \hz (-1 + \hz + \hx)},\nn\\[5pt]
&&\hat\sigma^{\it 9}_{DF1}=\frac{1}{N^2-1}\frac{\qt \hx}{2Q^2(-1 + \hz)},\nn\\[5pt]
&&\hat\sigma^{\it 9}_{GF1}=-\frac{1}{N^2-1}\frac{\qt \hx}{2Q^2(-1 + \hz)},\nn\\[5pt]
&&\hat\sigma^{{\it 9}}_{2}=-{\qt (-1 + 2 \hx)\over Q^2},\nn\\[5pt]
&&\hat\sigma^{{\it 9}}_{3}={\qt (-1 + 2 \hx)\over Q^2}.
\end{eqnarray}
This completes the specification of all partonic hard cross sections in (\ref{qFraXsecfinal}) and (\ref{twist2Xsec}).  

\section{Summary}

In this paper, we have studied the twist-3 cross section
for the transversely polarized hyperon production in SIDIS $ep\to e\Lambda^\uparrow X$
in the framework of collinear factorization. 
The cross section consists of five structure functions with different dependences on the azimuthal angles.  
We have presented the LO cross section which occurs from
the twist-3 DF in the initial proton combined with the transversity FF for $\Lambda^\uparrow$
and the twist-3 quark FFs for $\Lambda^\uparrow$ combined with the unpolarized DF in the proton
for all five structure functions.   The derived cross section is relevant for the large-$P_T$
$\Lambda^\uparrow$ production in the future EIC experiment.  
For completeness the contribution from the twist-3 purely gluon FFs
as well as another $q\bar{q}g$-FF of the type $\sim \la 0|gF^{\alpha w}|hX\ra \la hX| \bar{\psi} \psi|0\ra$ 
needs to be included (as discussed at the end of section 2), which
will be reported in a separate paper\cite{ikarashi2022}.

\section*{Acknowledgments}

This work has been supported by the Grant-in-Aid for
Scientific Research from the Japanese Society of Promotion of Science
under Contract Nos.~19K03843 (Y.K.) and 18J11148 (K.Y.),
National Natural Science Foundation in China 
under grant No. 11950410495, Guangdong Natural Science Foundation under
No. 2020A1515010794
and research startup funding at South China
Normal University.


\end{document}